\documentclass[11pt,twoside]{article}
\usepackage{asp2014}

\aspcpryear{2015}
\aspvoltitle{Proceedings of ASTRONUM-2014, 22nd-27th June 2014, Long Beach CA, USA}
\aspvolauthor{N.~Pogorelov and E.~Audit, eds.}
\resetcounters

\begin{document}

\title{The marriage of gas and dust}
\author{Daniel~J.~Price,$^1$ and Guillaume~Laibe,$^2$
\affil{$^1$Monash Centre for Astrophysics, Monash University, Vic, 3800, Australia; \email{daniel.price@monash.edu}}
\affil{$^2$School of Physics and Astronomy, University of St. Andrews, North Haugh, St. Andrews, Fife KY16 9SS, UK; \email{guillaume.laibe@gmail.com}}
}

\paperauthor{Daniel~J.~Price}{daniel.price@monash.edu}{0000-0002-4716-4235}{Monash University}{School of Mathematical Sciences}{Clayton}{Vic}{3800}{Australia}
\paperauthor{Guillaume~Laibe}{guillaume.laibe@gmail.com}{ORCID_Or_Blank}{University of St. Andrews}{School of Physics and Astronomy}{St. Andrews}{Fife}{KF16 9SS}{UK}

\begin{abstract}
Dust-gas mixtures are the simplest example of a two fluid mixture. We show that when simulating such mixtures with particles or with particles coupled to grids a problem arises due to the need to resolve a very small length scale when the coupling is strong. Since this is occurs in the limit when the fluids are well coupled, we show how the dust-gas equations can be reformulated to describe a single fluid mixture. The equations are similar to the usual fluid equations supplemented by a diffusion equation for the dust-to-gas ratio or alternatively the dust fraction. This solves a number of numerical problems as well as making the physics clear.
\end{abstract}

\section{Introduction}
 Dust holds the key to star and planet formation. It provides the main source of opacity in interstellar clouds, absorbing radiation at ultraviolet and optical wavelengths and re-emitting in the infrared. Dust is volatile, sublimating once $T \gtrsim 1000K$, providing a sensitive tracer of the star formation process. Dust grains grow to form planets.

From the perspective of this conference, dust-gas mixtures are interesting as the simplest example of a multi-fluid system where the two species are modelled as separate fluids coupled by a drag term, giving insight into more complicated systems such as non-ideal magnetohydrodynamics. In the case of dust and gas the system is given by
\begin{align}
\frac{\partial \rho_{\rm g}}{\partial t} + \nabla\cdot (\rho_{\rm g} {\bf v}_{\rm g} ) & = 0, \label{eq:ctyg}\\
\frac{\partial \rho_{\rm d}}{\partial t} + \nabla\cdot (\rho_{\rm d} {\bf v}_{\rm d} ) & = 0, \\
\frac{\partial {\bf v}_{\rm g}}{\partial t} + ( {\bf v}_{\rm g}\cdot\nabla) {\bf v}_{\rm g} & = -\frac{\nabla P_{\rm g}}{\rho_{\rm g}} + K \left( {\bf v}_{\rm d} - {\bf v}_{\rm g}\right) + {\bf f}, \\
\frac{\partial {\bf v}_{\rm d}}{\partial t} + ( {\bf v}_{\rm d}\cdot\nabla) {\bf v}_{\rm d} & = - K \left( {\bf v}_{\rm d} - {\bf v}_{\rm g}\right) + {\bf f}, \label{eq:momd}
\end{align}
where $\rho_{\rm g}$ and $\rho_{\rm d}$ are the gas and dust densities, respectively, ${\bf v}_{\rm g}$ and ${\bf v}_{\rm d}$ are the gas and dust velocities, $P_{\rm g}$ is the gas pressure, ${\bf f}$ represents external forces such as gravity and $K$ is a term hiding all the details of the drag interaction. This equation set has been widely used to model dust using both smoothed particle hydrodynamics, where both gas and dust are modelled using particles \citep[e.g.][]{monaghankocharyan95,mhm03,barriere-fouchetetal05,chanayakshin11,ayliffeetal12}, and in hybrid codes where the hydrodynamics is solved on a grid and the dust is simulated using particles \citep[e.g.][]{fromangpapaloizou06,paardekoopermellema06,youdinjohansen07,miniati10,baistone10}.

\articlefigure[width=0.45\textwidth]{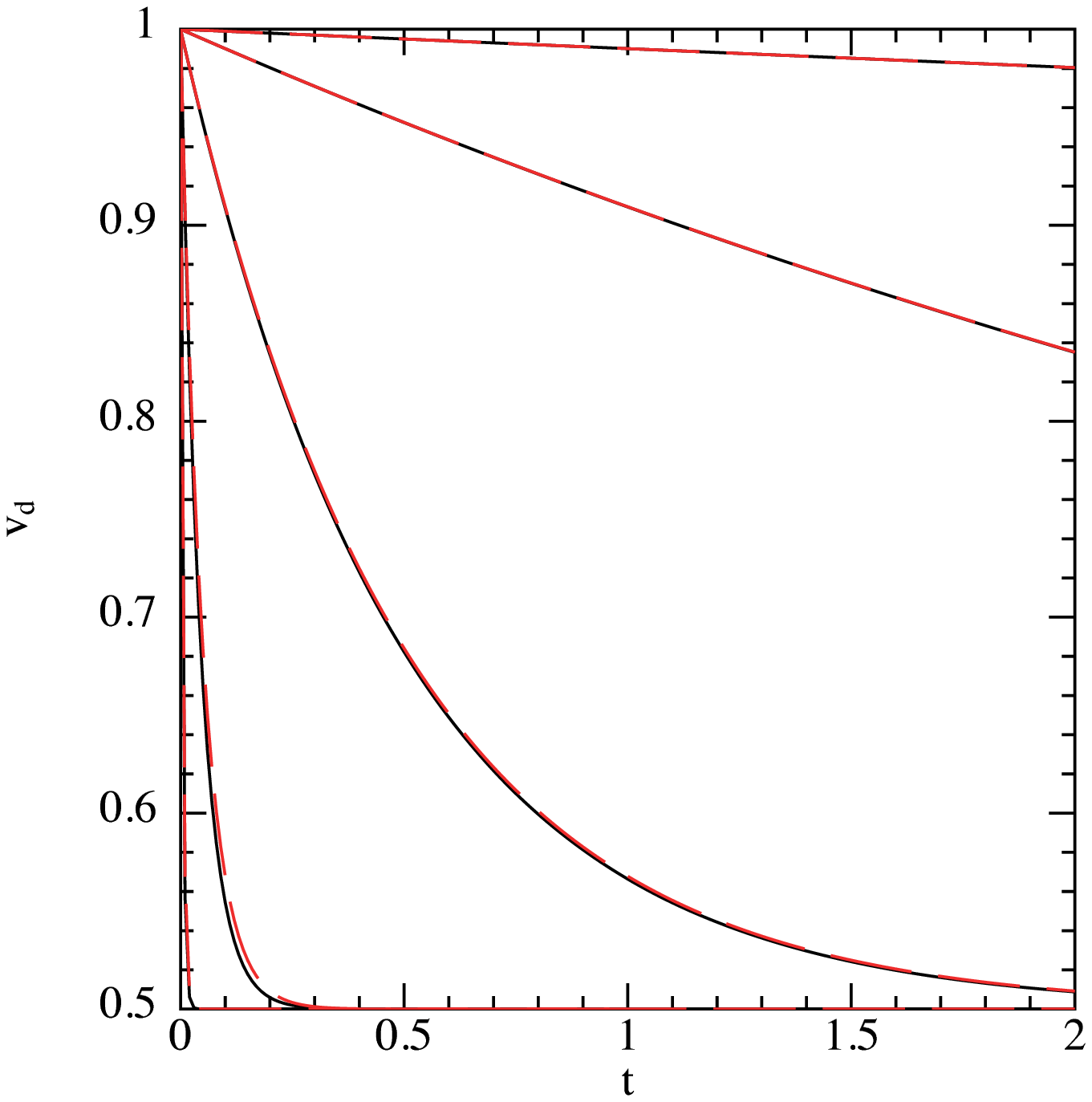}{fig:tstop}{The stopping time is the characteristic time for the decay of the relative velocity to the barycentre caused by drag. We show the decay of dust velocity as a function of time in a periodic one dimensional box containing gas and dust, where the dust is moving relative to the gas. Here we compare numerical results (red) with the analytic solution (black) for drag strengths varying between $K=0.01$ and $K=100$. High drag means short stopping times, which occurs for small grains.}

\subsection{Dust: a real drag}
The main physics from this set of equations that is different to the dynamics of each fluid separately is captured in the concept of the `stopping time', defined by
\begin{equation}
t_{\rm stop} \equiv \frac{\rho_{\rm d}\rho_{\rm g}}{K ( \rho_{\rm d} + \rho_{\rm g} )},
\end{equation}
which is inversely proportional to the drag. This is illustrated in Figure~\ref{fig:tstop} which shows numerical results compared to the analytic solution of perhaps the simplest test problem --- the `\textsc{dustybox}' \citep{monaghankocharyan95,paardekoopermellema06,laibeprice11}, involving dust initially moving at constant velocity relative to a uniform density gas at rest. The solution is an exponentially decaying relative velocity on a timescale determined by the stopping time. High drag results in short stopping times (strong coupling), while low drag results in long stopping times (weak coupling).

\subsection{Give us a wave}
A key problem we found when developing our numerical code was that, apart from the \textsc{dustybox}, there were no other simple problems with analytic solutions for dust-gas mixtures that could be used as a benchmark. Most codes in astrophysics have used the linear modes of the `streaming instability' \citep{youdingoodman05} but this requires a shearing box, and a perturbation to an already complicated equilibrium state. The physical meaning of the instability is also obscure, so it is hard to derive intuition from.

\articlefiguretwo{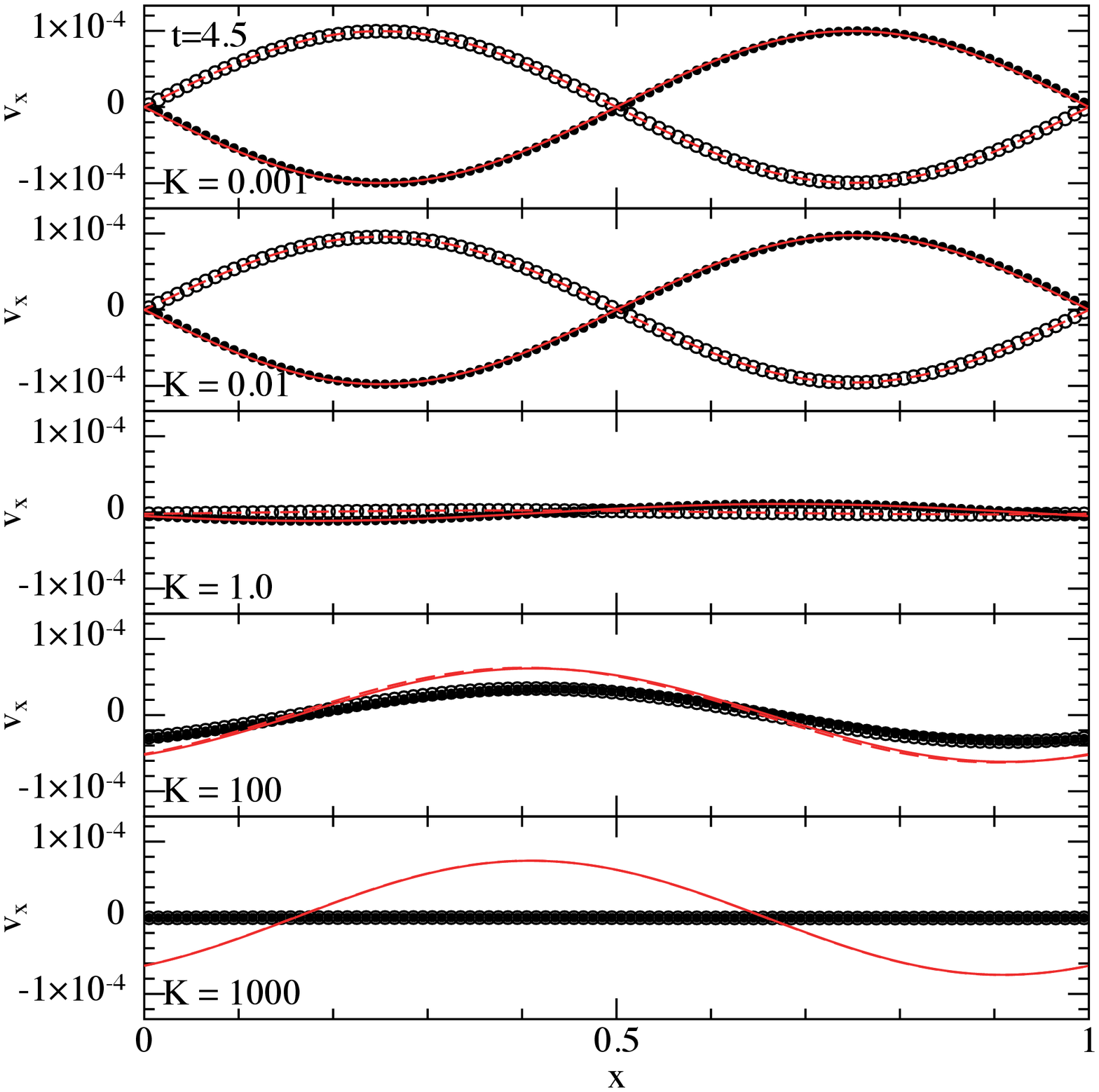}{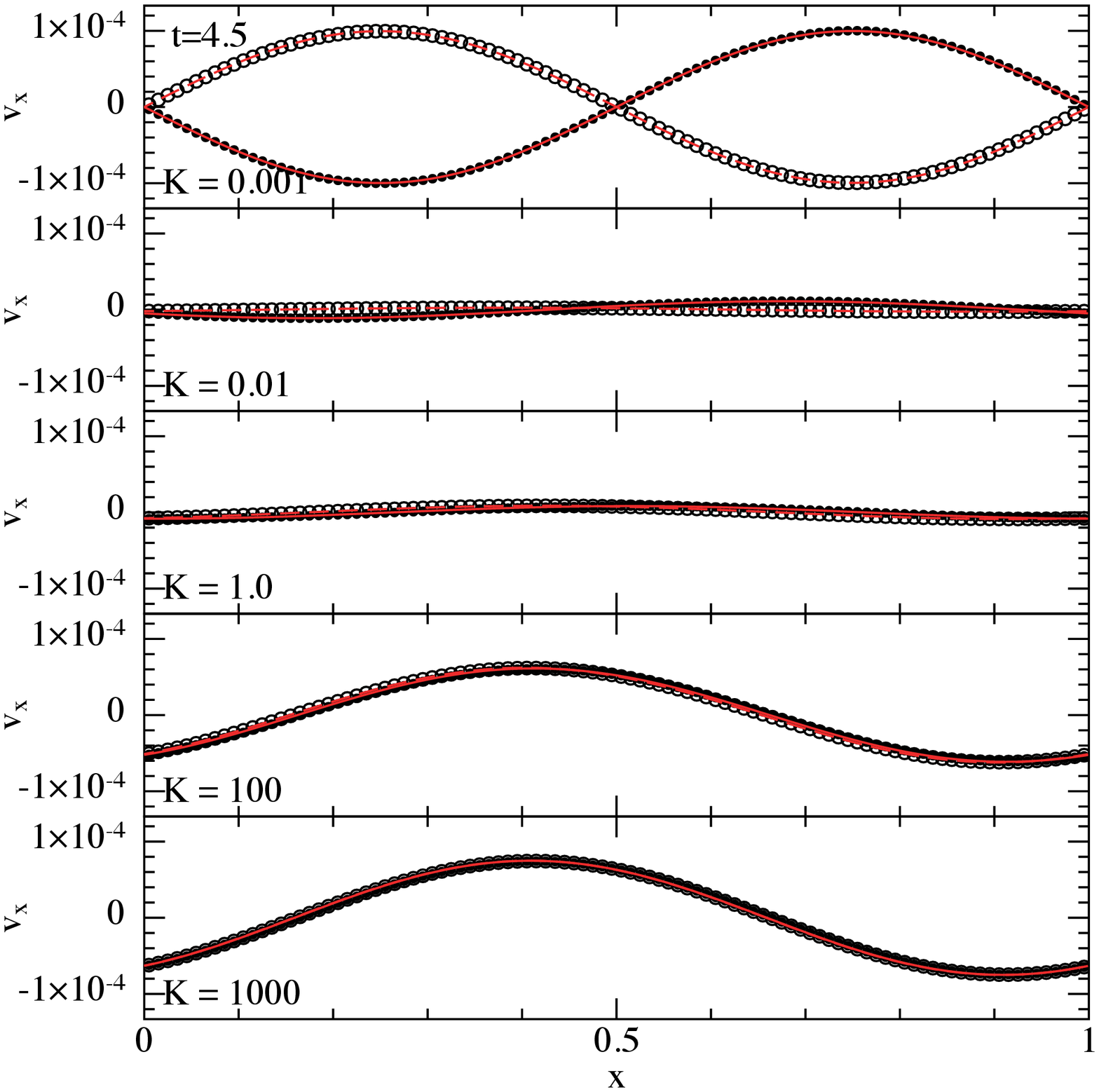}{fig:dustywave}{Propagation of waves in a dust-gas mixture, showing dust and gas velocities for the two-fluid (left) compared to a one-fluid (right) numerical solution (black solid and open circles for gas and dust, respectively), with both compared to the analytic solution (red solid/dashed lines for gas/dust). For low drag/long stopping times (top) the solution is an undamped sound wave, and the same is true at high drag/short stopping times apart from a phase shift (bottom). The drag terms dissipate energy most effectively at intermediate coupling. The two fluid approach (left) leads to overdamping in the limit of strong coupling (lower panels in left Figure), which is fixed using a one fluid approach, without loss of accuracy at low drag (right panels).}


 As one of us (GL) particularly enjoyed long derivations, we decided to fix this by deriving the analytic solution for linear waves in a dusty gas. We published this in \citet{laibeprice11}, although the eventual expressions were too long to write in the paper, so instead we provided some routines (attached to the arXiv version) to evaluate the solution given the relevant parameters. The solutions (red lines in Fig.~\ref{fig:dustywave}) proved extremely useful and provide a great deal of insight into the dynamics of the mixture.
 
  In hindsight the results are rather obvious --- at low drag/weak coupling/long stopping times the solution is an undamped sound wave in the gas, since the gas could not care less about the dust. It is less intuitive but also obvious in hindsight that this is also the solution at high drag/strong coupling/short stopping times (bottom panel in Fig.~\ref{fig:dustywave}). In this case the gas and dust are perfectly stuck to each other, so the only change is that the fluid is heavier, leading to an undamped wave propagating at the `modified sound speed'  \citep[e.g.][]{miuraglass82}
 \begin{equation}
 \tilde{c}_{\rm s} = c_{\rm s}\left(1 + \frac{\rho_{\rm g}}{\rho_{\rm d}}\right)^{-\frac12}.
 \end{equation}
 \emph{Only at intermediate stopping times is the solution strongly damped by the drag}.
 
 \subsection{Knowing the answer spoils the fun}
 The bottom panels on the left side of Figure~\ref{fig:dustywave} reveal a a deep and fundamental problem. This is that with a two fluid approach it is very difficult to converge on the analytic solution when the drag is strong. Without the analytic solution this would be easily dismissed --- ``of course drag is dissipative so the solution should be damped'' (though a careful numericist would \emph{always} do a resolution study, I hear you say).
 
 Importantly, this is a \emph{spatial} resolution issue, quite separate to the already well-known issue of needing small timesteps to ensure $\Delta t \lesssim t_{\rm stop}$. The reason for the non-convergent behaviour is intuitive from the \textsc{dustywave} solution, and relates to the length scales in the problem. At low drag or long stopping times the separation between the gas and dust is large. This separation is caused by the pressure gradient acting on the gas but not the dust. The drag force acts across this separation to pull the two fluids back together. During the ``bringing back together'' the relative kinetic energy is dissipated. When the drag is strong the pressure gradient causes a tiny separation, and the drag should act over a very short lengthscale. However, if the resolution length in the simulation is longer than this it will appear to be acting over a much larger lengthscale (the resolution length) and hence cause much more dissipation than should physically be present, similar to the behaviour of the fluid at intermediate drag. The same issue occurs for shocks \citep{laibeprice14a}.

\subsection{We just need an infinite supercomputer}
 In \citet{laibeprice12} we showed that the resolution criterion is
\begin{equation}
\Delta x \lesssim t_{\rm stop} c_{\rm s}, 
\end{equation}
as well as the usual $\Delta t \lesssim t_{\rm stop}$. While the timestepping issue may be solved with implicit methods \citep[e.g.][]{monaghan97a,laibeprice12a}, the spatial issue is much more difficult to fix. Worse, the limit $t_{\rm stop} \to 0$ ($K \to \infty$) implies $\Delta t \to 0$ and $\Delta x \to 0$, i.e. we require both an infinite number of timesteps \emph{and} infinite spatial resolution in the limit where the stopping time tends to zero. Worse still, this is the obvious limit where the fluids are perfectly coupled and waves propagate as undamped sound waves at the modified sound speed. Since we do not have access to infinite supercomputing capabilities a better approach is needed.

\section{Dusty gas with one fluid}
 The key \citep{laibeprice14} is that the problem occurs in the limit where the fluids are well coupled. In this case we can simply reformulate the equations to describe a single fluid moving with the velocity of the barycentre of both fluids, i.e.
\begin{equation}
{\bf v} \equiv \frac{\rho_{\rm g}{\bf v}_{\rm g} + \rho_{\rm d}{\bf v}_{\rm d}}{\rho_{\rm g} + \rho_{\rm d}}.
\end{equation}
Then, with a simple change of variables from ${\bf v}_{\rm g}$, ${\bf v}_{\rm g}$, $\rho_{\rm g}$ and $\rho_{\rm d}$ to ${\bf v}$, ${\bf v}_{\rm dr} \equiv {\bf v}_{\rm d} - {\bf v}_{\rm g}$, $\rho \equiv \rho_{\rm g} + \rho_{\rm d}$ and $\rho_{\rm d}/\rho_{\rm g}$ we can rewrite equations (\ref{eq:ctyg})--(\ref{eq:momd}) without loss of generality as
\begin{align}
\frac{{\rm d}\rho}{{\rm d}t} & = -\rho (\nabla\cdot{\bf v}), \\
\frac{{\rm d}{\bf v}}{{\rm d}t} & = -\frac{\nabla P_{\rm g}}{\rho} - \frac{1}{\rho}\nabla\cdot\left(\frac{\rho_{\rm g}\rho_{\rm d}}{\rho} {\bf v}_{\rm dr} {\bf v}_{\rm dr}\right) + {\bf f}, \label{eq:pr} \\
\frac{{\rm d}}{{\rm d}t} \left( \frac{\rho_{\rm d}}{\rho_{\rm g}}\right) & = - \frac{\rho}{\rho_{\rm g}^{2}} \nabla\cdot \left(\frac{\rho_{\rm g}\rho_{\rm d}}{\rho} {\bf v}_{\rm dr}\right), \label{eq:dtg}\\
\frac{{\rm d} {\bf v}_{\rm dr}}{{\rm d}t} & = -\frac{{\bf v}_{\rm dr}}{t_{\rm stop}} + \frac{\nabla P_{\rm g}}{\rho_{\rm g}} - ({\bf v}_{\rm dr}\cdot\nabla) {\bf v} + \frac12 \nabla \left[\frac{\rho_{\rm d} - \rho_{\rm g}}{\rho_{\rm g} + \rho_{\rm d}} {\bf v}_{\rm dr}^{2}\right]. \label{eq:vdiff}
\end{align}
The physics is clear when presented this way: The first term on the right hand side of the equation for the drift velocity (\ref{eq:vdiff}) results in exponential decay of the differential velocity on the stopping time. The second term, the pressure gradient, causes the differential velocity to grow. The effect of the drift velocity is to change the dust-to-gas ratio via (\ref{eq:dtg}). Better still, the first two equations are just the usual equations of hydrodynamics, with modifications to the equations of motion (\ref{eq:pr}) because the pressure depends on the gas pressure, not the total pressure, and due to an anisotropic pressure term. It is also clearly evident that in the limit ${\bf v}_{\rm dr} \to 0$ the equations reduce to the usual equations of single-fluid gas dynamics, and solves the spatial resolution issue (Figure~\ref{fig:dustywave} demonstrates this). As ${\bf v}_{\rm dr}$ still changes on the stopping time, we still require implicit time integration when $t_{\rm stop}$ is short, but this is easily achieved\footnote{An open-source implementation of all of our one-fluid and two-fluid algorithms in smoothed particle hydrodynamics are available from \url{http://users.monash.edu.au/~dprice/ndspmhd}} \citep{laibeprice14a}.

\subsection{The diffusion approximation for dust}
The limit can be made even clearer, and the equations simpler, by making the `terminal velocity approximation' in which the drift velocity is assumed to have reached its asymptotic value \citep[e.g.][]{youdinjohansen07,laibeprice14}
\begin{equation}
{\bf v}_{\rm dr} \approx t_{\rm stop} \frac{\nabla P_{\rm g}}{\rho_{\rm g}},
\end{equation}
which is a valid approximation when the stopping time is shorter than any other timescale in the calculation, i.e. for $t_{\rm stop} < \Delta t$. In this case the equations are \citep{laibeprice14}
\begin{align}
\frac{{\rm d}\rho}{{\rm d}t} & = -\rho (\nabla\cdot{\bf v}), \\
\frac{{\rm d}{\bf v}}{{\rm d}t} & = -\frac{\nabla P_{\rm g}}{\rho} + {\bf f}, \\
\frac{{\rm d}\epsilon}{{\rm d}t} & = - \frac{1}{\rho} \nabla\cdot \left(\epsilon t_{\rm stop} \nabla P\right), \label{eq:dtgdiff}
\end{align}
where we have also used the dust fraction $\epsilon \equiv \rho_{\rm d}/\rho$ instead of the dust-to-gas ratio. Hence we can simulate dust-gas mixtures with only two simple modifications to the equations of hydrodynamics: pressure depending on gas density not total density, and a diffusion equation for the dust fraction. This equation can be evolved explicitly for small grains since the diffusion coefficient is small when $t_{\rm stop}\to 0$, and can be formulated in the code to exactly conserve the mass of both species. Interestingly, diffusion would control the timestep precisely where the approximation itself breaks down, which seems to be a general truth (i.e. that where diffusion controls the timestep is where diffusion itself is a bad approximation to the physics; think radiation).

\subsection{The observers know something after all}
In the limit of ${\bf v}_{\rm dr} = 0$ we obtain a constant dust-to-gas ratio, and the only modification to hydrodynamics is to the sound speed. A constant dust-to-gas ratio is commonly assumed when inferring gas properties from extinction maps of dust in molecular clouds. Hence we provide a rigorous mathematical justification for this, although to what extent this approximation holds in the interstellar medium is an open question.

\section{Summary}
We have derived a new single-fluid formulation for dust-gas mixtures. This solves both the spatial and temporal resolution problems for small grains. In addition we have shown that this equation set can be simplified further to provide a `diffusion approximation for dust' that consists of the usual fluid equations supplemented by a diffusion equation for the dust fraction. The net result is a method analogous to existing methods for treating non-ideal MHD, e.g. where ion-neutral drift is represented as a diffusion term in the induction equation. We recently generalised this to handle multiple dust species \citep{laibeprice14b}. Applications to star and planet formation are underway.

\acknowledgements We thank Joe Monaghan and Ben Ayliffe for insightful discussions. This work was funded by an Australian Research Council (ARC) Discovery Project Grant (DP1094585) and a Future Fellowship (FT130100034).

\bibliography{dan}

\end{document}